\documentclass[conference]{IEEEtran}
\usepackage{amsmath}
\usepackage{amssymb}
\usepackage[dvips]{graphicx}
\usepackage{epsfig}

\newtheorem{lemma:bitenergylow}{Lemma}
\newtheorem{lemma:bitenergyhigh}[lemma:bitenergylow]{Lemma}

\newtheorem{prop:asympcap}{Theorem}
\newtheorem{prop:flashminbitenergy}[prop:asympcap]{Theorem}
\newtheorem{prop:flashbitenergy}[prop:asympcap]{Theorem}
\newtheorem{prop:pasympcap}[prop:asympcap]{Theorem}

\begin{document}


\title{ Energy-Efficient Modulation Design for Reliable Communication in Wireless Networks }



%
\author{\authorblockN{Qing Chen and Mustafa Cenk Gursoy}
\authorblockA{Department of Electrical Engineering\\
University of Nebraska-Lincoln, Lincoln, NE 68588\\ Email:
chenqing@huskers.unl.edu, gursoy@engr.unl.edu}}


\maketitle

\begin{abstract}
In this paper, we have considered the optimization of the $M$-ary quadrature amplitude modulation (MQAM)
constellation size to minimize the bit energy consumption under
average bit error rate (BER) constraints. In the computation of the energy
expenditure, the circuit, transmission, and retransmission energies
are taken into account. A combined log-normal shadowing and Rayleigh
fading model is employed to model the wireless channel. The link
reliabilities and retransmission probabilities are determined
through the outage probabilities under log-normal shadowing effects.
Both single-hop and multi-hop transmissions are considered.  Through
numerical results, the optimal constellation sizes are identified.
Several interesting observations with practical implications are
made. It is seen that while large constellations are preferred at
small transmission distances, constellation size should be decreased
as the distance increases. Similar trends are observed in both fixed
and variable transmit power scenarios. We have noted that variable
power schemes can attain higher energy-efficiencies. The analysis of
energy-efficient modulation design is also conducted in multi-hop
linear networks. In this case, the modulation size and routing paths
are jointly optimized, and the analysis of both the bit energy and
delay experienced in the linear network is provided.
%
\end{abstract}

\section{Introduction}

In wireless communications, the analysis of energy efficiency and
reliable packet transmission has attracted much interest recently,
spurred by emerging wireless applications, such as in wireless
sensor networks, where recharging and/or replacing the batteries of
the wireless devices are difficult. In such scenarios, a wireless
node can transmit and receive a finite number of bits before its
battery runs out. Therefore, minimizing the bit-energy consumption
will be a key factor in maximizing the node and, eventually, the
network lifetime.

In wireless networks, operation at all levels of the communication
protocol stack has an impact on the energy consumption and
therefore, energy efficiency has to be addressed in device,
physical, link, and network layers jointly. In early work in the
networking literature, minimum-energy routing algorithms, which
select the paths with the minimum total transmission power over all
the nodes, are developed (see e.g., \cite{chang}, \cite{scott}). As
described in \cite{SBAM}, if the links are assumed to be error-free
and hence there is no need for retransmission, energy-efficient
routing algorithms choose the minimum-hop paths in cases in which
the transmitter power is fixed. On the other hand, in
variable-transmission power scenarios, the sender nodes dynamically
vary their transmission power $P_{t}$ proportional to $d^\beta$,
where $d$ is the link distance and $\beta$ is the path loss
exponent, so that the received signal power is kept fixed to achieve
an acceptable signal-to-noise ratio (SNR) level at the receiver.  In
such cases, again when error-free links are considered,
energy-efficient routing protocols tend to choose routes with a
large number of small-range hops. We note that the assumption of
error-free links has led to the design of algorithms
which consider the energy spent only in a single transmission over each link.

However, in practical wireless channels, transmitted signals are
subject to random propagation effects such as shadowing and
multipath fading, leading to random fluctuations in the strength of
the received signal. As a result of these random variations,
reliable reception of the transmitted message signal cannot be
guaranteed all the time. In cases in which the received signal power
is below a certain threshold required for acceptable performance,
outage is said to have occurred and retransmission of the
transmitted signal is required. Due to their significant impact on
the overall energy consumption, retransmissions, which are generally
handled by the link layer, must be considered in the analysis of
energy-efficient
operation. 
Two types of retransmission schemes, namely end-to-end
retransmission (\emph{EER}) and hop-by-hop retransmission
(\emph{HHR}), are elaborated in \cite{SBAM} and \cite{JZXW}, where
the relationship between the link error probabilty of individual
links and the overall number of retransmissions needed to ensure
reliable packet delivery is established. In \cite{SBAM}, the authors
designed energy-efficient routing algorithms in the network layer by
also taking into account the link layer retransmission probabilities
and energies.

Clearly, design in the device and physical layers has a significant
impact on the energy expenditure. For instance, transmission power
depends on the selected modulation type and size, required error
probability levels, and assumed channel attenuation models, all of
which are mainly physical layer considerations. At the device level,
besides the transmission energy, the circuit energy of the
microchips performing as radio frequency transceivers, A/D and D/A
converters and baseband processors or other application interfaces
in the battery-driving nodes is an additional source of energy
consumption. The circuit energy, together with the transmission
energy, has to be considered in the overall optimization modeling,
especially in short range range networks in which the circuit energy
is comparable to the transmission energy,  and also as the number of
retransmissions increases.

So far, several studies have addressed different aspects of
energy-efficient operation. Optimization of the one-hop transmission
distance without retransmission mechanism is studied in
\cite{PCBODEC} while optimal routing with retransmission mechanism
considering \emph{EER} and \emph{HHR} models is investigated
\cite{SBAM}. In these studies, physical and device layer
considerations are not incorporated in the models in detail. Authors
in \cite{SCAJG1} and \cite{SCAJG2} consider both circuit and
transmission energy consumption and address the optimization of
modulation size under energy constraints. However, in these studies,
link layer outage events, link reliabilities, and retransmission
energies are not considered in the optimization.

In this paper, we provide a holistic approach by studying the energy
efficiency considering device, physical, link, and network layers
jointly. In particular, we investigate the optimal size of the
quadrature amplitude modulation (QAM) constellation that minimizes
the energy required to send one bit of information. An accurate
energy consumption model is used, in which the total energy equals
to the sum of circuit energy (device layer), transmission energy
(physical layer), retransmission energy (link layer). Initial
results are obtained for single-hop transmissions. Subsequently,
results are extended to a simple linear multi-hop network model for
which energy-efficient routing paths (network layer) are studied.

The remainder of the paper is organized as follows. Section II
introduces the log-normal shadowing effects and link outage
probability. Section III presents the accurate energy consumption
model and discusses how to compute the average receive power $\overline{P}_{r}$ over
the Rayleigh fading channel given an average BER constraint
$\overline{p}_{b}$ in uncoded square MQAM. Section IV presents the
linear network model and analyzes the bit energy consumption, delay
performance, the optimal routing for a specific MQAM constellation
and also for the optimal MQAM constellation.

\section{Log-Normal Shadowing and Link Reliability}
\label{sec:shadowing}
A combined path loss and log-normal shadowing model is considered as
a model for the large-scale propagation effects. This
has been confirmed empirically to accurately model the variations in
the received power in indoor and outdoor radio propagation
environments \cite{AD}-\cite{VE}. For this model, the ratio of
the received power to transmission power in dB is given by
\begin{align}
\frac{P_{r}}{P_{t}}(dB)=K_{dB}-10\beta\log_{10}\frac{d}{d_{0}}-\psi_{dB}
\end{align}
where $\psi_{dB}$ is a Gaussian-distributed random variable with
zero mean and variance $\sigma_{\psi_{dB}}^{2}$, modeling random
shadowing effects; $d_{0}$ is a reference distance for the antenna
far-field; $\beta$ is the path loss exponent; $K_{dB} =
20\log_{10}\frac{\lambda}{4\pi d_{0}} $ is a constant that depends on
the wave length $\lambda$ of the transmitted signal and $d_0$. In
\cite{JKAN}, an approximation for the packet error probability has
been presented for the log-normal shadowing model. We hereby adopt
the same idea to define the outage probability under path loss and
shadowing as the probability that $P_{r}$ falls below a threshold
power $P_{min}$. This outage probability of a certain link is
denoted by $p_{link}$ and is given by
\begin{align}
\hspace{-.19cm}p_{link}&=p(P_{r}(d)\leq P_{min})\\
&=1-Q\Big(\frac{P_{min}-(P_{t}+K_{dB}-10\beta\log_{10}(d/d_{0}))}{\sigma_{\psi_{dB}}}\Big)
\label{eq:outageprob}
\end{align}
where $Q(\cdot)$ is the Gaussian $Q$-function \cite{AD}. We assume
that if outage occurs and hence $P_{r}(d)\leq P_{min}$, then
receiver doesn't successfully receive the packet and the sender
needs to
retransmit the packet until successful reception is achieved. 
In our analysis, 
we adopt the hop-by-hop retransmission (\emph{HHR}) scheme. We
employ $p_{link}$ to calculate the average number of retransmissions
in a statistical sense. If we denote $E_{ij}$ as the energy
consumption of single transmission over a link between node $i$ and
node $j$ with link error probability $p_{link_{ij}}$, we can derive
the average total energy consumption for a reliable packet delivery
over this link in the \emph{HHR} mode as
\begin{align}
E_{ij}^{total}=\frac{E_{ij}}{1-p_{link_{ij}}}.
\end{align}
To compute this total energy $E_{ij}^{total}$ and the link error
probability $p_{link_{ij}}$, we have to specify the parameters in
(\ref{eq:outageprob}). Generally, we assume $d_{0}=1$ and employ the
empirical results from \cite{SSG} for ${\sigma^{2}_{\psi_{dB}}}$ and
$\beta$. In the following section, we describe how the threshold
power level $P_{min}$ is specified.

\section{Accurate Energy Consumption Model for BER Constrained MQAM}

Usually in energy-constrained networks, the emphasis on minimizing
the transmission energy is reasonable in long-range networks
$(d\geq100m)$, where the transmission energy is dominant in the
total energy consumption. On the other hand, in many recently
proposed densely distributed networks such as wireless sensor
networks, the average distance between nodes could be within 10 to
100 meters, for which the circuit energy consumption is comparable
to the transmission energy. In such cases, the circuit energy
consumption should be included in the total energy consumption for
more accurate analysis.
\subsection{Circuit and Transmission Energies}

We adopt the accurate energy consumption formulation from
\cite{SCAJG2}, \cite{SCAJG3} and \cite{SCAJG4}. We assume that the
source node has $L$ bits in one packet to transmit. One single
transmission is composed of three distinct periods: transmission
period $T_{on}$, transient period $T_{tr}$, and sleeping period
$T_{sp}$. Accordingly, the total energy required to send $L$ bits is
represented by
\begin{align}
E=P_{on}T_{on}+P_{sp}T_{sp}+P_{tr}T_{tr} \intertext{where}
 P_{on}=P_{t}+\alpha
P_{t}+P_{ct}+P_{cr}.
\end{align}
Above $P_{on}$, $P_{sp}$ and $P_{tr}$ are power consumptions for the
active mode, sleep mode and transient mode, respectively.
Specifically, $P_{on}$ comprises the transmission power $P_{t}$, the
amplifier power $\alpha P_{t}$, the circuit power at the transmitter
$P_{ct}$ and the circuit power at the receiver $P_{cr}$. We assume
that the power of the sleep mode is very small and we approximate
it as $P_{sp} \approx 0$. Now, the bit energy is
\begin{align}
E_{a}=(((1+\alpha)P_{t}+P_{ct}+P_{cr})T_{on}+P_{tr}T_{tr})/L.
\end{align}
When we consider uncoded MQAM as our modulation scheme, we can write
\begin{align} \label{eq:Ton}
T_{on}=\frac{LT_{s}}{b}=\frac{L}{bB}
\end{align}
where $b$ is the constellation size defined as $b=\log_{2}M$ in
MQAM, and $T_s$ is the symbol duration. Note that in the second
equality in (\ref{eq:Ton}),  the approximation $T_{s}\approx1/B$ is
used. Therefore, in a specific MQAM, if the number of bits $L$ and
channel bandwidth $B$ are given, the active period $T_{on}$ is
consequently derived. Finally, we note that the amplifier efficiency
for MQAM can be obtained from $\alpha=\frac{\xi}{\eta}-1$ where $\xi=3\frac{\sqrt{M}-1}{\sqrt{M}+1}$ is a function of the MQAM constellation size.

\subsection{Average BER Constraint for MQAM}

We consider a wireless channel model in which, in addition to the
large-scale log-normal shadowing effects discussed in Section
\ref{sec:shadowing}, small-scale Rayleigh fading is experienced.
First, we note that for square-constellation MQAM, the symbol error
probability $P_{s}$ as a function of the instantaneous received SNR
$\gamma_{s}$ is given by \cite{AD}:
\begin{multline}
P_{s}=\frac{4(\sqrt{M}-1)}{\sqrt{M}}Q\left(\sqrt{\frac{3\gamma_{s}}{M-1}}\right)\\
-\frac{4(M-2\sqrt{M}+1)}{M} Q^{2}
    \left(\sqrt{\frac{3\gamma_{s}}{M-1}}\right) \label{eq:insterrorprob}
\end{multline}
where the Gaussian $Q$-function and its square can be expressed,
respectively, as
\begin{align}
Q(z)=\frac{1}{\pi}\int_{0}^{\pi/2}
\exp\left[\frac{-z^2}{2\sin^2\phi}d\phi\right] \label{eq:qfunc}
\\
Q^2(z)=\frac{1}{\pi}\int_{0}^{\pi/4}
\exp\left[\frac{-z^2}{2\sin^2\phi}d\phi\right]. \label{eq:q2func}
\end{align}
The average symbol error probability is computed by averaging the
instantaneous error probability over the specific fading
distribution $p_{\gamma_{s}}(\gamma)$ as follows
\begin{align}
\overline{P}_{s}=\int_{0}^\infty{P_{s}(\gamma)p_{\gamma_{s}}(\gamma)d\gamma}. \label{eq:avgerrorprob}
\end{align}
Substituting (\ref{eq:insterrorprob})-(\ref{eq:q2func})
into (\ref{eq:avgerrorprob}),
we can express the average BER for Rayleigh fading approximately as
\begin{multline}
\overline{p}_{b}=\frac{4}{\pi\log_{2}M}\left(1-\frac{1}{\sqrt{M}}\right)\int_{0}^{\pi/2}M_{\gamma_{s}}\left(-\frac{g}{\sin^2\phi}\right)d\phi
\\-\frac{4}{\pi\log_{2}M}\left(1-\frac{1}{\sqrt{M}}\right)^2\int_{0}^{\pi/4}M_{\gamma_{s}}\left(-\frac{g}{\sin^2\phi}\right)d\phi\\
=\frac{4}{\pi\log_{2}M}\left(1-\frac{1}{\sqrt{M}}\right)\int_{0}^{\pi/2}\left(1+\frac{3{\overline{\gamma}}_{b}\log_{2}M}{2(M-1)\sin^2\phi}\right)^{-1}d\phi\\
\frac{4}{\pi\log_{2}M}\left(1-\frac{1}{\sqrt{M}}\right)^2\int_{0}^{\pi/4}\left(1+\frac{3\overline{\gamma}_{b}\log_{2}M}{2(M-1)\sin^2\phi}\right)^{-1}d\phi.
\end{multline}
Above, $M_{\gamma}(s)$ denotes the moment generating function
\begin{align}
M_{\gamma}(s)=\int_{0}^\infty p_{\gamma}(\gamma)e^{s\gamma}d\gamma
\end{align}
of the fading distribution $p_{\gamma_s}(\gamma)$ in the Rayleigh fading channel where
\begin{align}
p_{\gamma_{s}}(\gamma)=\frac{1}{\overline{\gamma}_{s}}e^{-\gamma/\overline{\gamma}_{s}}.
\end{align}
For a given BER constraint $\overline{p}_{b}$, we can use (13) to
compute the required bit SNR $\overline{\gamma}_{b}$ numerically. Then,
the minimum average received power that guarantees an average BER of
$\overline{p}_{b}$ is computed by \cite{AD}:
\begin{align}
P_{min}=\overline{\gamma}_{b}N_{0}B\log_{2}M
\end{align}
where $\frac{N_{0}}{2}$ is the noise power spectrum density. Below
this power level, the received information packets might undergo
unacceptable distortion and the transmission is considered
unreliable, which is quantitatively modeled by the outage
probability in the log-normal shadowing model in Section
\ref{sec:shadowing}.

\section{Optimal Modulation for Single-Hop and Multi-Hop Transmissions}

In this section, we numerically evaluate the optimal QAM
constellation size that minimizes the bit energy consumption under
different BER constraints, and establish the tradeoffs associated
with this analysis. The bit energy is computed by taking into
account circuit, transmission, and retransmission energies.
Initially, we consider single-hop transmissions and then investigate
multi-hop links.

\subsection{Optimal MQAM Constellation Size in Single-Hop Transmissions}

In this section, we study the bit energy consumption in single-hop
transmissions where retransmission, circuit energy, uncoded square
MQAM and $\overline{p}_{b}$ constraint are considered based on
equations (3), (4), (7), (13), and (16). We consider two transmit
power policies: fixed $P_{t}$ and variable $P_{t}$. 


\begin{table}
\caption{Network and Circuit Parameters} \label{table:minbitenergy}
\begin{center}
\begin{tabular}{|c|c|}

\hline $\overline{p}_{b}=0.0001$ &$\eta$=0.35
\\
\hline $N_{0}=4*10^{-21}W/Hz$ & $\beta$=3.12
\\
\hline $L$=20000 bits & $freq=2.5*10^{9}Hz$
\\
\hline $P_{tr}$=100mW & $P_{ct}$=98.2mW
\\
 \hline $\sigma_{\psi_{dB}}$=3.8dB & $P_{cr}$=112.5mW
\\
\hline $B$=10KHz & $T_{tr}$=5$\mu{s}$
\\
\hline

\end{tabular}
\end{center}
\end{table}

The values of the set of parameters used in numerical results are
provided in Table 1. For instance, the maximum allowable average BER is $\overline{p}_{b}=0.0001$. In fixed transmit power scheme, we assume
$P_{t}$=100mW and vary the constellation size $b\in[2,4,6,8,10]$ at
link distance $d\in[5m,25m,50m,75m,100m]$. The bit energy
consumption vs. constellation size curves for different transmission
distances are given in Figure 1.
\begin{figure}
\begin{center}
\includegraphics[width = 0.45\textwidth]{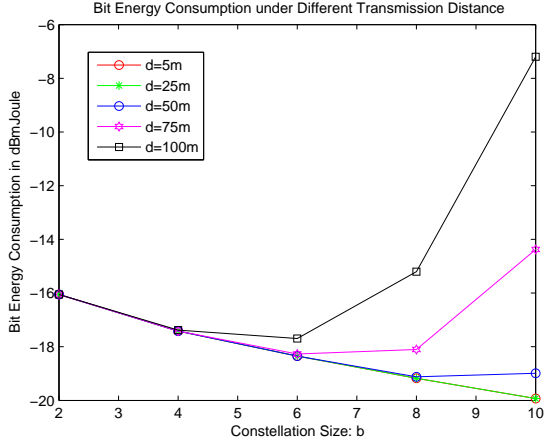}
\caption{Bit-energy consumption vs. constellation size $b$ when
$P_t$ is fixed} \label{fig:1}
\end{center}
\end{figure}
We immediately observe that the bit energy increases as the
constellation size gets either large or very small, and there exists
an optimal $b$ value at which the bit energy is minimized. This tradeoff is due to the following. We note that an MQAM with large constellation sends the information at a faster rate decreasing $T_{on}$ and hence the transmit energy. However, a large constellation requires  a larger $P_{min}$, leading to a higher outage probability and consequently more retransmissions. On the other hand, a small constellation size requires less retransmissions but increases $T_{on}$ and hence the transmit energy. Hence, the optimal constellation size should provide a balance between these effects. In Fig. 1,  we observe that the optimal $b$ is 8 at $d=50m$ and the optimal $b$ is 6
at $d=75m$. We note that the energy-minimizing constellation
size gets smaller with increasing distance. In Fig. 2., the bit
energy is plotted as a function of transmission distance for
different constellation sizes. We see that while large constellation
sizes are performing well at small distances, small constellations
should be preferred when the distance gets large.
\begin{figure}
\begin{center}
\includegraphics[width = 0.45\textwidth]{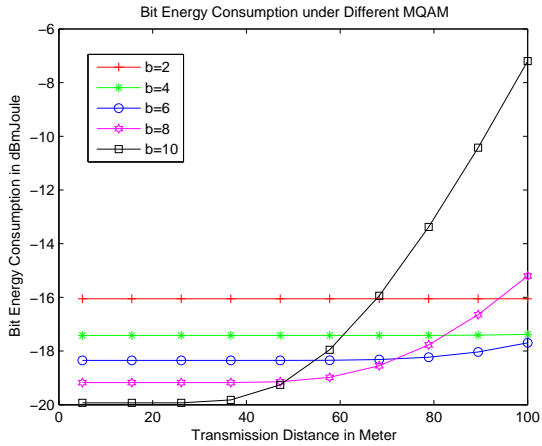}
\caption{Bit-energy consumption vs. transmission distance when $P_t$
is fixed} \label{fig:1}
\end{center}
\end{figure}

Now, we consider a variable transmit power scheme, where $P_{t}$ is
dynamically adjusted so that the average received power is
\begin{equation}
P_{min} = P_{t}+K_{dB}-10\beta\log_{10}(d/d_{0}).
\end{equation}
In this case, it can be easily seen that the link error probability
$p_{link}$ in (3) is always equal to 0.5 independent of the transmission
distance. We apply the same configuration to compute the bit energy
for different MQAM constellation sizes and transmission distance.
Figures 3 and 4 provide the numerical results. Conclusions similar
to those for the fixed-transmit power case can be drawn in the
variable-power case as well. However, we note that smaller bit
energies and hence higher energy efficiency can be attained in
the variable-power case as evidenced in the numerical results.
\begin{figure}
\begin{center}
\includegraphics[width = 0.45\textwidth]{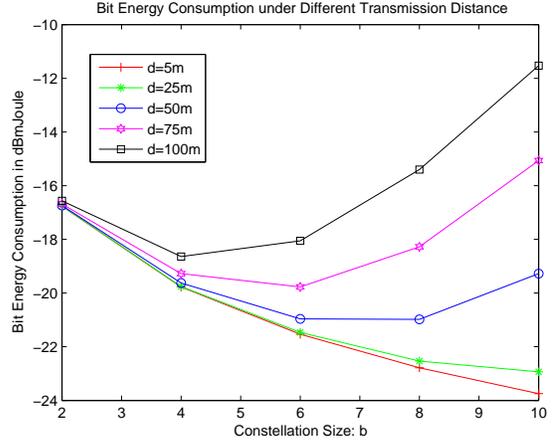}
\caption{Bit-energy consumption vs. constellation size $b$ when
$P_t$ is variable } \label{fig:1}
\end{center}
\end{figure}
\begin{figure}
\begin{center}
\includegraphics[width = 0.45\textwidth]{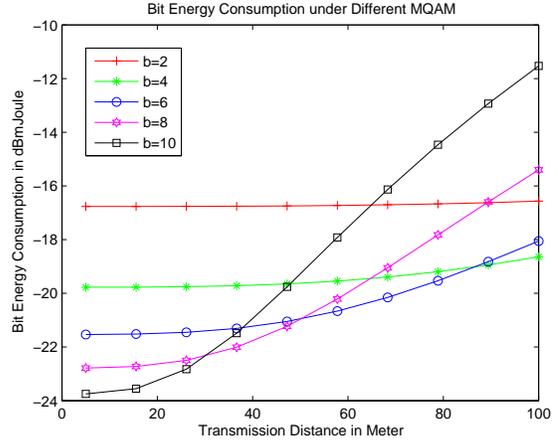}
\caption{Bit-energy consumption vs. transmission distance when $P_t$
is variable } \label{fig:1}
\end{center}
\end{figure}

\subsection{Linear Network Model }

In this section, we study multi-hop scenarios and  consider a simple
linear network with $N+2$ nodes: source, destination and $N$
intermediate nodes equally distributed in between. Each intermediate
node can behave as either an active relay or just a sleeping node.
The working condition of $N$ intermediate nodes could be visualized
by a binary code, where the active nodes are denoted as 1 and the
sleeping nodes as 0. In this model, totally $2^{N}$ routing paths
are available in the \emph{HHR} relay transmission scenario.
Accordingly, we can either choose a routing path with less
intermediate relay nodes and longer hops, or a routing path with
more intermediate relay nodes and shorter hops. Note that longer
hops tend to have higher $p_{link}$, resulting in more
retransmissions while more intermediate relay nodes imply a
potentially higher circuit energy consumption. Therefore, the selection of the optimal route should
strike a balance between these tradeoffs. 

\begin{figure}
\begin{center}
\includegraphics[width = 0.40\textwidth]{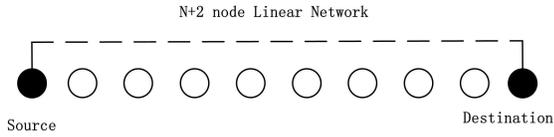}
\caption{Topology of Linear Network} \label{fig:1}
\end{center}
\end{figure}
In the linear network, we set the distance between the source and destination to 100m, consider 9 intermediate
nodes, and choose $\overline{p}_{b}$ from the following set: \{0.0001,
0.0003, 0.0005, 0.0008, 0.001\}. We apply the parameters in Table 1 and employ
the method used in Section IV.A for each possible relay link. Given the constellation size $b = \log_2 M$
and a specific $\overline{p}_{b}$, we search the $2^{N}$ routing
paths to find the optimal route that requires the minimum bit energy consumption defined as:
\begin{align}
 minimize ({E_{a}}_{total}^{HHR})=\min{\sum_{i,j}\frac{Ea_{ij}}{1-p_{link_{ij}}}}.
\end{align}
At fixed $P_{t}$=100mW scheme, among all possible routings, the
optimal routing is found in terms of the minimum bit energy
consumption when the packet containing 20000 bits is relayed in
\emph{HHR} mode successfully from the source to the destination via
intermediate nodes. Fig. 6 shows the minimum bit energy consumption
corresponding to the optimal routing for each square uncoded MQAM
under different BER constraints when $P_{t}$ is fixed. Hence, in the figure, the bit energies for different constellation sizes are the ones required by the optimal route. Again, there exists a certain constellation size that minimizes the bit energy consumption. Moreover, we observe that lower error probabilities expectedly require more energy. Fig. 7 shows
the corresponding delay experienced in optimal routing. The delay is
formulated as the sum of time intervals consumed over the relay
links, whose individual delay is computed as:
\begin{align}
Delay_{link_{ij}}=\frac{T_{on}+T_{r}}{1-p_{link_{ij}}}
\end{align}
Fig. 7 is showing that on points which achieve optimal bit energy
consumption, the corresponding delay is also minimal. This is due to the fact that the optimal route and modulation size that minimize the energy requirements are favorable in terms of delay as well, because they try to diminish both $T_{on}$ and the number of retransmissions by finding a balance between competing factors in the fixed-transmit-power case.

\begin{figure}
\begin{center}
\includegraphics[width = 0.40\textwidth]{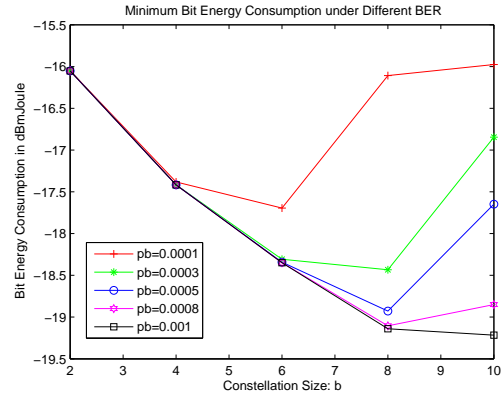}
\caption{Optimal bit energy consumption vs. constellation size $b$
when $P_t$ is fixed} \label{fig:1}
\end{center}
\end{figure}

\begin{figure}
\begin{center}
\includegraphics[width = 0.40\textwidth]{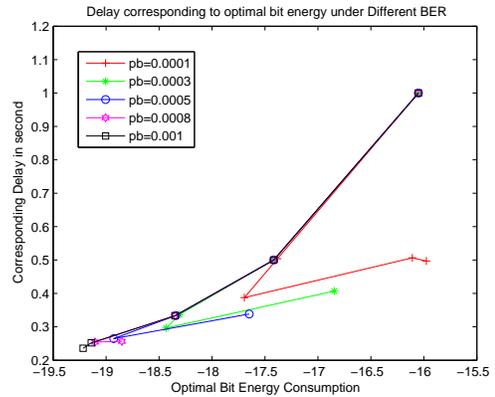}
\caption{Delay vs. optimal bit energy consumption when $P_t$ is
fixed} \label{fig:1}
\end{center}
\end{figure}

Finally we consider the variable $P_{t}$ scheme, and adopt all
previous parameters to repeat the bit energy as well as delay
analysis. Figure 8 shows the minimum bit energy consumption
corresponding to the optimal routing for each square uncoded MQAM
when $P_{t}$ is variable. Compared to Fig. 6, the variable $P_{t}$
scheme shows better performance, requiring
about 1-1.5 dBmJoule less than that of the fixed $P_{t}$ scheme at
optimal MQAM points. Fig. 9 plots the optimal bit energy-delay
curves. Here, we note that it no longer holds that optimal bit energy
and optimal delay are achieved at the same point for all BER
constraints $\overline{p}_{b}$.
\begin{figure}
\begin{center}
\includegraphics[width = 0.40\textwidth]{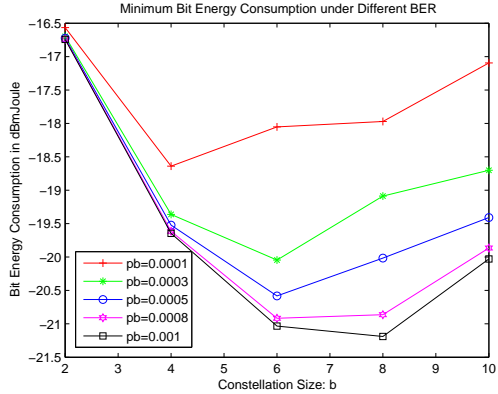}
\caption{Optimal bit energy consumption vs. constellation size $b$
when $P_t$ is variable} \label{fig:1}
\end{center}
\end{figure}
\begin{figure}
\begin{center}
\includegraphics[width = 0.40\textwidth]{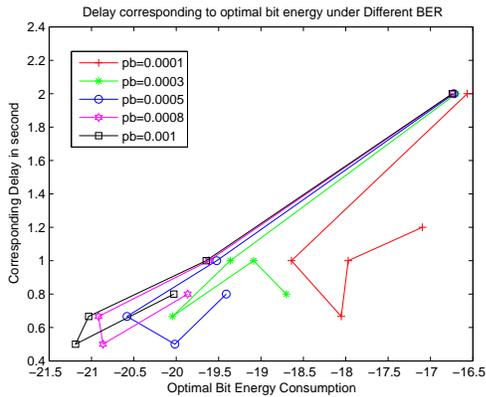}
\caption{Optimal bit energy consumption vs. constellation $b$ when
$P_t$ is variable} \label{fig:1}
\end{center}
\end{figure}

Intuitively, in the fixed $P_{t}$ scheme, the transmit power $P_{t}$
would strongly affect $p_{link_{ij}}$. Accordingly, we can also vary
$P_{t}$ as well as constellation size $b$ to find the global minimum
point under a given $\overline{p}_{b}$ constraint. Fig. 10
illustrates optimal bit energy consumption corresponding to the
optimal routing as a function of $b$ and $P_{t}$. The numerical
result shows the global minimum is -19.71 $dBmJoule$ for the error rate constraint $\overline{p}_b$=0.0001, found when
$b$=4 and $P_{t}$=25mW. 

\begin{figure}
\begin{center}
\includegraphics[width = 0.40\textwidth]{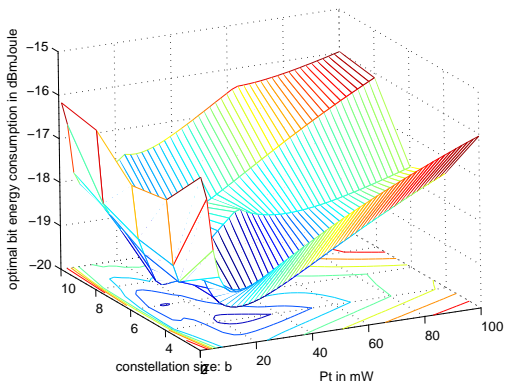}
\caption{Optimal bit energy consumption vs. constellation size $b$
and transmit power $P_t$.} \label{fig:1}
\end{center}
\end{figure}

\section{Conclusion}

In this paper, we have considered the optimization of the MQAM
constellation size to minimize the bit energy consumption under
average BER constraints. In the computation of the energy
expenditure, the circuit, transmission, and retransmission energies
are taken into account. A combined log-normal shadowing and Rayleigh
fading model is employed to model the wireless channel. The link
reliabilities and retransmission probabilities are determined
through the outage probabilities under log-normal shadowing effects.
Both single-hop and multi-hop transmissions are considered.  Through
numerical results, the optimal constellation sizes are identified.
Several interesting observations with practical implications are
made. It is seen that while large constellations are preferred at
small transmission distances, constellation size should be decreased
as the distance increases. Similar trends are observed in both fixed
and variable transmit power scenarios. We have noted that variable
power schemes can attain higher energy-efficiencies. The analysis of
energy-efficient modulation design is also conducted in multi-hop
linear networks. In this case, the modulation size and routing paths
are jointly optimized, and the analysis of both the bit energy and
delay experienced in the linear network is provided.



\begin{thebibliography}{99}

\bibitem{chang} J.-H. Chang and L. Tassiulas, ``Energy conserving routing in wireless ad-hoc networks," Proc. of INFOCOM, March 2000.

\bibitem{scott} K. Scott and N. Bamboos, ``Routing and channel assignment for low power transmission in PCS," Proc. of ICUPC, Oct. 1996.

%

\bibitem{SBAM}Suman Banerjee, Archan Misra, ``Minimum Energy Paths for
Reliable Communication in Multi-hop Wireless networks,'' MOBIHOC,
pp.146-156, June. 2002.


\bibitem{JZXW}Jinhua Zhu, Xin Wang,``On Accurate Energy Consumption
Models for Wireless Ad Hoc Networks,'' IEEE Trans. Wireless Commun.,
vol.5, no.11, pp.3077-3086, Nov. 2006.

\bibitem{PCBODEC}Priscilla Chen, Bob O'Dea, Ed Callaway
``Energy Efficient System Design with Optimum Transmission Range for
Wireless Ad Hoc Networks,'' Proc. of ICC 2002, vol. 2, pp. 945- 952.

\bibitem{SCAJG1}Shuguang Cui, Andrea J. Godsmith, Ahmad Bahai,
``Energy-constrained Modulation Optimization,'' IEEE Trans. Wireless
 Commun.,vol.4 no.5 pp.2349-2360, Sep. 2005.

 \bibitem{SCAJG2}Shuguang Cui, Andrea J. Goldsmith, Ahmad Bahai,
``Energy-Efficiency of MIMO and Cooperative MIMO Techniques in
Sensor Networks,'' IEEE \emph{J.Select. Areas Commun,} vol. 32, no.
6, pp.1089-1098, Aug. 2004.

\bibitem{JKAN}Johnson Kuruvila, Amiya Nayak, Ivan Stojmenovic,
``Hop Count Optimal Position-Based Packet Routing Algorithms for Ad
Hoc Wireless Networks With a Realistic Physical Layer,'' IEEE
\emph{J.Select. Areas Commun,} vol.23, no.6, Jun. 2006.

\bibitem{AD}Andrea J. Goldsmith
``Wireless Communications", Cambridge, UK: Cambridge University Press,
2005.


\bibitem{SSG}S.S. Ghassemzadeh, L.J. Greenstein, A. Kavcic, T.
Sveinsson, V. Tarokh ``UWB Indoor Path Loss Model for Residential
and Commercial Buildings,'' Proc. of VTC 2003, pp. 3115- 3119 Vol.5, Oct 2003.

\bibitem{VE}V.Erceg, L.J. Greenstein, S.Y. Tjandra, S.R. Parkoff, A.Gupta, B. Kulic,
A.A. Julius, and R. Bianchi, ``An Emperically Based Path Loss Model
for Wireless Channels in Suburban Environments,'' IEEE
\emph{J.Select. Areas Commun,} vol. 17, no. 7, pp.1205-1211, Jul.
1999.

\bibitem{SCAJG3}Shuguang Cui, Andrea J. Goldsmith,
``Energy Efficient Routing Based on Cooperative MIMO Techniques,'' Proc. of ICASSP, pp.805-808, Mar. 2005.

\bibitem{SCAJG4}Shuguang Cui, Andrea J. Goldsmith, Sanjay Lall,
``Joint Routing, MAC, and Link Layer Optimization in Sensor Networks
with Energy Constraints,'' Proc. of ICC 2005, vol. 2, pp.725 - 729, May. 2005.

\bibitem{DC}John. G. Proakis, Masoud Salehi,
``Fundamentals of Communication Systems'', Prentice Hall, Dec. 2004.










\end{thebibliography}
\end{document}